\pdfoutput=1
\let\mypdfximage\pdfximage
\protected\def\pdfximage{\immediate\mypdfximage}

\documentclass[letterpaper,headsepline,DIV=15,BCOR=20mm]{scrartcl}

\usepackage[automark]{scrpage2}
\usepackage[USenglish]{babel}
\usepackage{graphicx,pstricks}
\usepackage{graphics}
\usepackage{epsfig}
\usepackage{subfig}
\usepackage{verbatim}
\usepackage{multirow}
\usepackage[pdftex=true,hyperindex=true,colorlinks=true,breaklinks=true, urlcolor=blue, linkcolor=black,citecolor=black,bookmarksnumbered=true,pdfpagelabels=true,plainpages=false]{hyperref}

\usepackage{booktabs} 
\ifpdf \usepackage{pdflscape} \else \usepackage{lscape} \fi 

\usepackage{savesym}
\usepackage[intlimits]{amsmath}
\savesymbol{iint}
\restoresymbol{TXF}{iint}
\usepackage{amssymb}
\numberwithin{equation}{section}
\numberwithin{figure}{section}
\numberwithin{table}{section}

\usepackage[sectionbib]{chapterbib} 

\usepackage[sectionbib,square,sort&compress]{natbib} 

\setcitestyle{numbers}
\setcitestyle{citesep={,}}  

\renewcommand\bibname{References} 

\usepackage{chngcntr}
\counterwithout{figure}{section}
\counterwithout{table}{section}

\newcommand{\Fig}[1]{Fig.~\ref{#1}}
\newcommand{\Figure}[1]{Figure~\ref{#1}}
\newcommand{\Tab}[1]{Tab.~\ref{#1}}
\newcommand{\Table}[1]{Table~\ref{#1}}

\newcommand{\unit}[1]{\,\ensuremath{\mathrm{#1}}}

\begin{document}
\renewcommand{\bibname}{References}

\title{The Cornell-BNL FFAG-ERL Test Accelerator}

\subtitle{White Paper}

\author{
\begin{large}
\begin{minipage}{0.8\textwidth}
\vspace{1cm}
Ivan Bazarov, John~Dobbins, Bruce Dunham, Georg~Hoffstaetter, Christopher~Mayes, Ritchie~Patterson, David~Sagan
\begin{center}
\emph{Cornell University, Ithaca NY}
\end{center}
\vspace{0.5cm}
Ilan Ben-Zvi, Scott Berg, Michael Blaskiewicz, Stephen Brooks, Kevin~Brown, Wolfram Fischer, Yue Hao, Wuzheng Meng, Fran\c{c}ois~M\'{e}ot, Michiko Minty, Stephen~Peggs, Vadim~Ptitsin, Thomas~Roser, Peter~Thieberger, Dejan Trbojevic, Nick Tsoupas.
\begin{center}
\emph{Brookhaven National Laboratory, Upton NY}
\end{center}
\begin{center}
\vspace{1cm}
\includegraphics[width=0.95\textwidth]{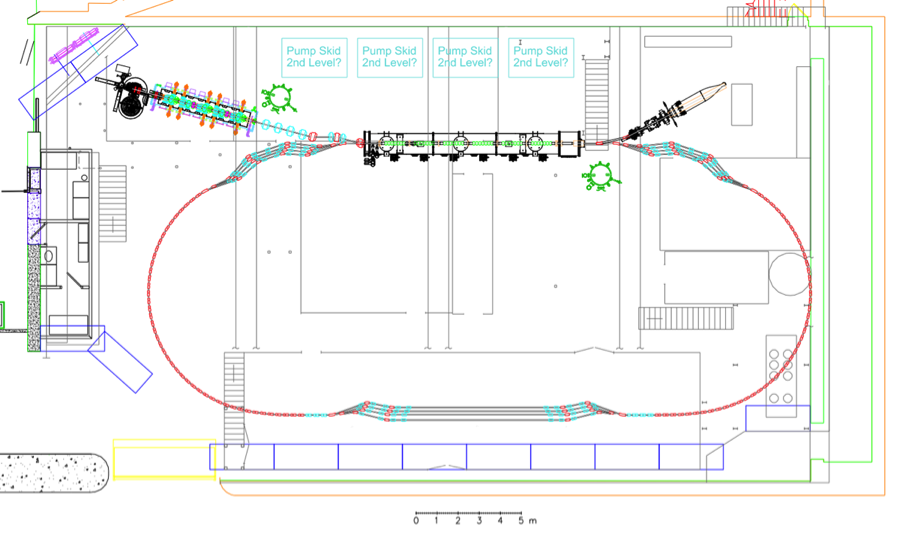}
\vspace{1cm}
\end{center}
\end{minipage}
\end{large}
}

\date{December 16, 2014}

%



\maketitle


%
\clearpage
\section{Executive Summary}

The Cornell-BNL FFAG-ERL Test Accelerator (C$\beta$) will be a unique resource to carry out accelerator science and enable exciting research in nuclear physics, materials science and industrial applications.

C$\beta$ will comprise the first ever Energy Recovery Linac (ERL) based on a Fixed Field Alternating Gradient (FFAG) lattice.  In particular, we plan to use a Non Scaling FFAG (NS-FFAG) lattice that is very compact and thus space- and cost- effective, enabling multiple passes of the electron beam in a single recirculation beam line, using the superconducting RF (SRF) linac multiple times.  The FFAG-ERL moves the cost optimized linac and recirculation lattice to a dramatically better optimum.

The prime accelerator science motivation for C$\beta$ is proving that the FFAG-ERL concept works.  This is an important milestone for the BNL plans to build a major Nuclear Physics facility, eRHIC, based on producing 21 GeV electron beams to collide with the RHIC ion beams~\cite{ref1}.  A consequence of the C$\beta$ work would be the availability of significantly better, cost-effective, compact CW high-brightness electron beams for a plethora of scientific investigations and applications, such as X-ray sources, dark-matter and dark-energy searches, industrial high-power Free-Electron Laser (FEL) applications, and much more.

C$\beta$ brings together the resources and expertise of a large DOE National Laboratory, BNL, and a leading research university, Cornell.  C$\beta$ will be built in an existing building at Cornell, for the most part using components that have been developed under previous R\&D programs, including a fully commissioned world-leading photoemission electron injector, a large SRF accelerator module, and a high-power beam stop.  The only elements that require design and construction from scratch is the FFAG magnet transport lattice.

This white paper describes a project that promises to propel high-power, high-brightness electron beam science and applications to an exciting new level.  The collaborative effort between Brookhaven and Cornell will be a model for future projects between universities and national lab, taking advantage of the expertise and resources of both to investigate new topics in a timely and cost-effective manner.

\clearpage
\section{Primary motivations} 
BNL is planning to transform RHIC to eRHIC by installing a new electron accelerator in the existing tunnel, providing polarized electrons with 3He ions, or with ions from deuterons to Uranium.  Significant simplification and cost reduction is possible by configuring eRHIC with non-scaling, Fixed-Field Alternating Gradient (NS-FFAG) optics in combination with an ERL.  Two NS-FFAG beamline arcs placed on top of each other allow multiple passes through a single superconducting linac.  Such magnets and optical techniques will be prototyped in the Cornell-BNL FFAG-ERL Test Accelerator (C$\beta$), along with other innovative accelerator technologies, commissioning eRHIC instrumentation and confirming the theories for multi-pass recirculative beam breakup.

The high beam power and high brightness provided by C$\beta$ will also enable exciting and important physics experiments including dark matter and dark energy searches~\cite{ref2}, Q-weak tests at lower energies~\cite{ref3}, proton charge radius measurements, and an array of polarized-electron enabled nuclear physics experiments. High brightness, narrow line width gamma rays can be generated by Compton scattering~\cite{ref4} using the ERL beam, and can be used for nuclear resonance fluorescence, the detection of special nuclear materials, and an array of astrophysical measurements.  The energy and current range of C$\beta$ will also be ideal for studying high power FEL physics for materials research and for industrial uses.

\section{The L0E experimental hall at Cornell}

\begin{figure}[tb]
\centering
\includegraphics[width=0.95\textwidth]{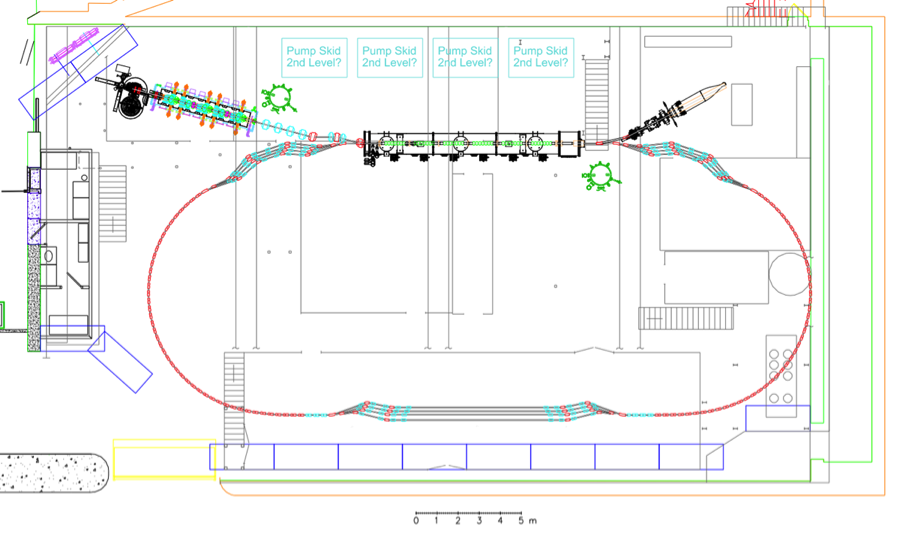}
\caption[]{Possible floor plan of the Cornell-BNL FFAG-ERL Test Accelerator in the L0E experimental hall in Wilson laboratory.}
\label{fig:diagram_cbeta_in_l0e}
\end{figure}

\Figure{fig:diagram_cbeta_in_l0e} shows the C$\beta$ lattice in the L0E experimental hall in Wilson laboratory at Cornell, while \Tab{tab:cbeta_parameters} lists the primary accelerator parameters.  \Figure{fig:diagram_cbeta_in_l0e} illustrates that there is sufficient space for all of the required hardware.  At the bottom and at the right of the accelerator are external walls.  The upper wall separates the area from the Cornell Electron Storage Ring (CESR), while on the left side are X-ray stations of the Cornell High Energy Synchrotron Source (CHESS).

\begin{table}[tb]
\caption[]{Primary parameters of the Cornell-BNL FFAG-ERL Test Accelerator.}
\begin{tabular*}{\columnwidth}{@{\extracolsep{\fill}}lll}
\toprule
Parameter & Value    & Unit     \\
\midrule
Linac energy gain	& 70 	& MeV \\
Injection energy	& 6 	& MeV \\
Linac passes	& 8 total: 4 accelerating + 4 decelerating	& \\
7 energy sequence	& 76, 146, 216, 286, 216, 146, 76 	& MeV \\
RF frequency	& 1300	& MHz \\
RF period		& 0.7692	& ns \\
Circumference harmonic	& 279	& \\
Accelerator circumference	&	64.34	& m \\
Revoultion period	& 0.2146	& \unit{\mu s} \\
Electron gun current 	& 100	& mA \\
Normalised RMS emittance & 2 (at 1~nC)	& \unit{\mu m} \\
Typical arc beta functions 	& 0.5 (ranges from 0.05 to 2.5)	& m \\
Typical RMS beam size	& 50 (ranges from 20 to 80)	& \unit{\mu m} \\
Bunch charge: commissioning& $ \approx 0.1$ & nC \\
\hphantom{Bunch charge:}     physics operation (CW)  &  0.1 to 1.0 & nC \\
\hphantom{Bunch charge:}     Dedicated experiments  &  $<5$ & nC \\

\bottomrule
\end{tabular*}
\label{tab:cbeta_parameters}
\end{table}

The most expensive C$\beta$ components --- the injector and the linac cryomodule --- are already built and are available for use at Cornell.  The L0E location is already served by significant accelerator infrastructure and other resources, including a liquid helium plant.  C$\beta$ is an important part of future plans at Cornell for accelerator research, nuclear physics research, materials studies, and ERL studies.

\clearpage
\section{NS-FFAG optics in an ERL}
A first NS-FFAG proof-of-principle Electron Model for Multiple Applications (EMMA) was built and operated at Daresbury Laboratory in 2012~\cite{ref5}.  The broader concept of an FFAG was revived recently, but is not new, originally developed by three independent groups in the 1950's~\cite{ref6,ref7, ref8}.  However, FFAG optics have not yet been used in combination with an ERL, a combination that is at the heart of the eRHIC design, and which is a core motivation for C$\beta$ accelerator physics prototyping.  In eRHIC the advantages of NS-FFAG optics are evident: many passes with a shorter superconducting linac and only two electron beamlines in each RHIC arc sector.  The C$\beta$ implementation will enable these prototype magnet and optical technologies to be studied.  Advanced understanding of NS-FFAG beam dynamics will also spin-off into related applications, such as hadron therapy, gantry optics, multi-pass neutron generators, and Accelerator Driven Subcritical Reactor (ADSR) systems.

The NS-FFAG concept was originally developed as a solution for fast acceleration of short lifetime muons.  Small dispersion functions lead to magnet apertures that are much smaller than with scaling FFAG optics~\cite{ref9}.  General advantages include:
\begin{enumerate}
  \item Energy range factors of 4 or 5 with constant, possibly permanent magnets.
  \item Small physical aperture, small magnet size, and small orbit offsets.
  \item Linear magnetic fields often enable the use of displaced quadrupoles.
  \item Large dynamic apertures, with no non-linear magnets or fields.
  \item Extreme focusing --- very small Twiss and dispersion functions.
\end{enumerate}
General disadvantages of NS-FFAG optics are:
\begin{enumerate}
  \item Tune and chromaticity variation with energy.
  \item Time-of-flight depends parabolically on beam energy deviation from the central reference energy.
\end{enumerate}

\clearpage
\section{Schedule}
Installation andy prototyping will be performed in several stages over two to three years, with incremental commissioning occurring as the hardware becomes available.   The major schedule stages --- with a critical path through magnet delivery --- are:
\begin{enumerate}
  \item  Clear out the experimental hall (3 months, year 1)
  \item  Install the injector and linac cryomodule (3 months, year 1)
  \item  Install cryogenics and RF power (2 months, year 1)
  \item Commission the injector and cryomodule (3-6 months, year 1)
  \item Commission single-turn FFAG-ERL (6 months, year 2)
  \item Install FFAG magnets and arcs (4 months, year 2)
  \item Commission FFAG return loops (year 3)
  \item  FFAG experiments (year 3)
\end{enumerate}
The injector has already been tested, but needs to be relocated to a new area, re-commissioned, and  The main linac cryomodule will not have undergone any testing before its installation.  The first commissioning step is to cool down the entire system to 2~K, taking from one week to one month.  Once cold, each cavity will be tested individually using a 5~kW solid state amplifier.  Multi-cavity tests will be performed when more amplifiers are available.  Cavity Q's will be measured, and the operation of the tuners, couplers and HOM's will be checked.  Testing will take 3 to 6 months.

\clearpage
\section{Optics and beam dynamics}

\begin{figure}[tb]
\centering
\includegraphics[width=0.95\textwidth]{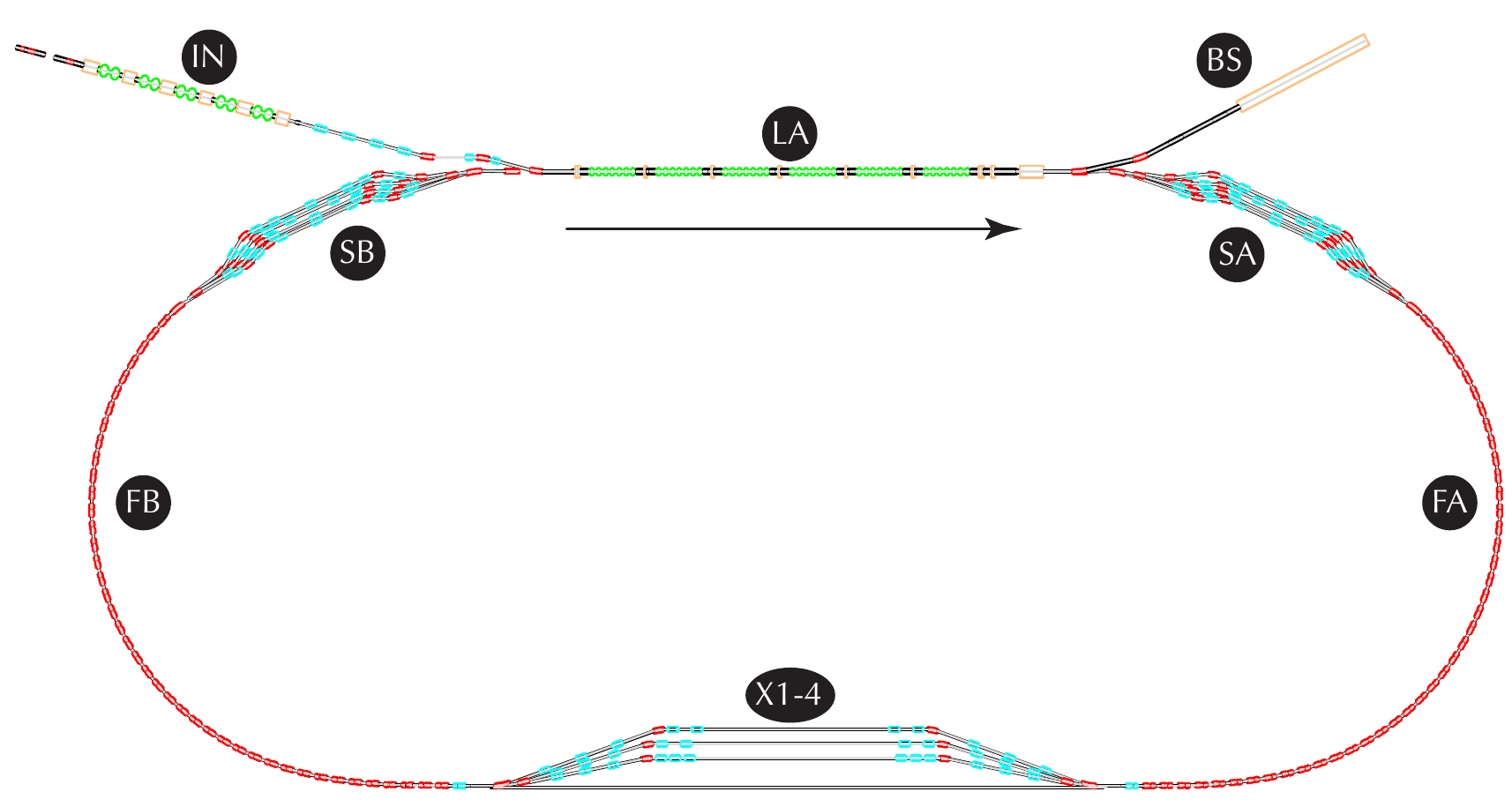}
\caption[]{Major optical components.  IN - Injector.  LA - linac.  SA, SB - splitter-combiner sections.  FA, FB - FFAG arcs.  X1-4 - experimental and diagnostic beamline sections.  BS - beam stop. }
\label{fig:layout_cbeta}
\end{figure}

\Figure{fig:layout_cbeta} shows the major optical components.  Starting from the injector (IN), the beam is merged into the linac (LA) at 6 MeV, and accelerated in the first pass by 70 MeV.  This lowest energy beam is separated from beams at other energies when entering the first splitter-combiner section (SA).  It merges back with the other beams to travel on the 76 MeV stable orbit in the FFAG arc (FA), before being separated again in the experimental/diagnostic beamline section (X1-4), merging into FFAG arc (FB), followed by a second splitter-combiner section (SB) for matching into another accelerating pass in the linac.  The process continues with 4 arc energies of 146 MeV, 216 MeV and 286 MeV, for a total of four accelerating passes.  The path length on the fourth pass is chosen so that re-entry into the linac is on the decelerating phase, and so that energy is recovered in the remaining four passes through the linac.  Finally the spent 6 MeV beam is directed to the beam stop (BS), which also includes 6D beam diagnostics.

The somewhat complicated splitter-combiner sections are necessary for orbit correction and to match the relatively large transverse beams in the   They also tune the net path lengths and momentum compaction factors to ensure efficient energy recovery.  The experimental area offers space for further optics tuning and diagnostics.   The arcs of the racetrack are made of doublet NS-FFAG cells, with a focusing quadrupole QF that is radially displaced by 5.4~mm, and a defocusing magnet BD that could be a displaced quadrupole, or a combined function magnet with only mid-plane symmetry.  

\begin{figure}[tb]
\centering
\subfloat[Orbit trajectories]{\label{figure_orbits_in_ffag_cell}\includegraphics[width=0.45\textwidth]{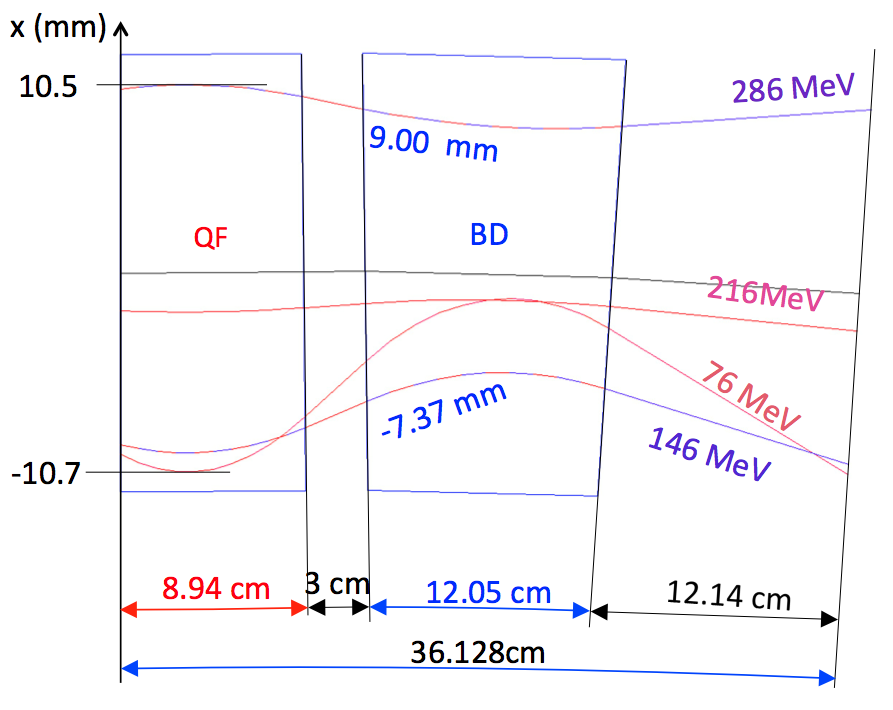}}
\hspace{0.05\textwidth}
\subfloat[Matched betatron and dispersion functions]{\label{plot_ffag_cell_beta_functions_with_energy}\includegraphics[width=0.45\textwidth]{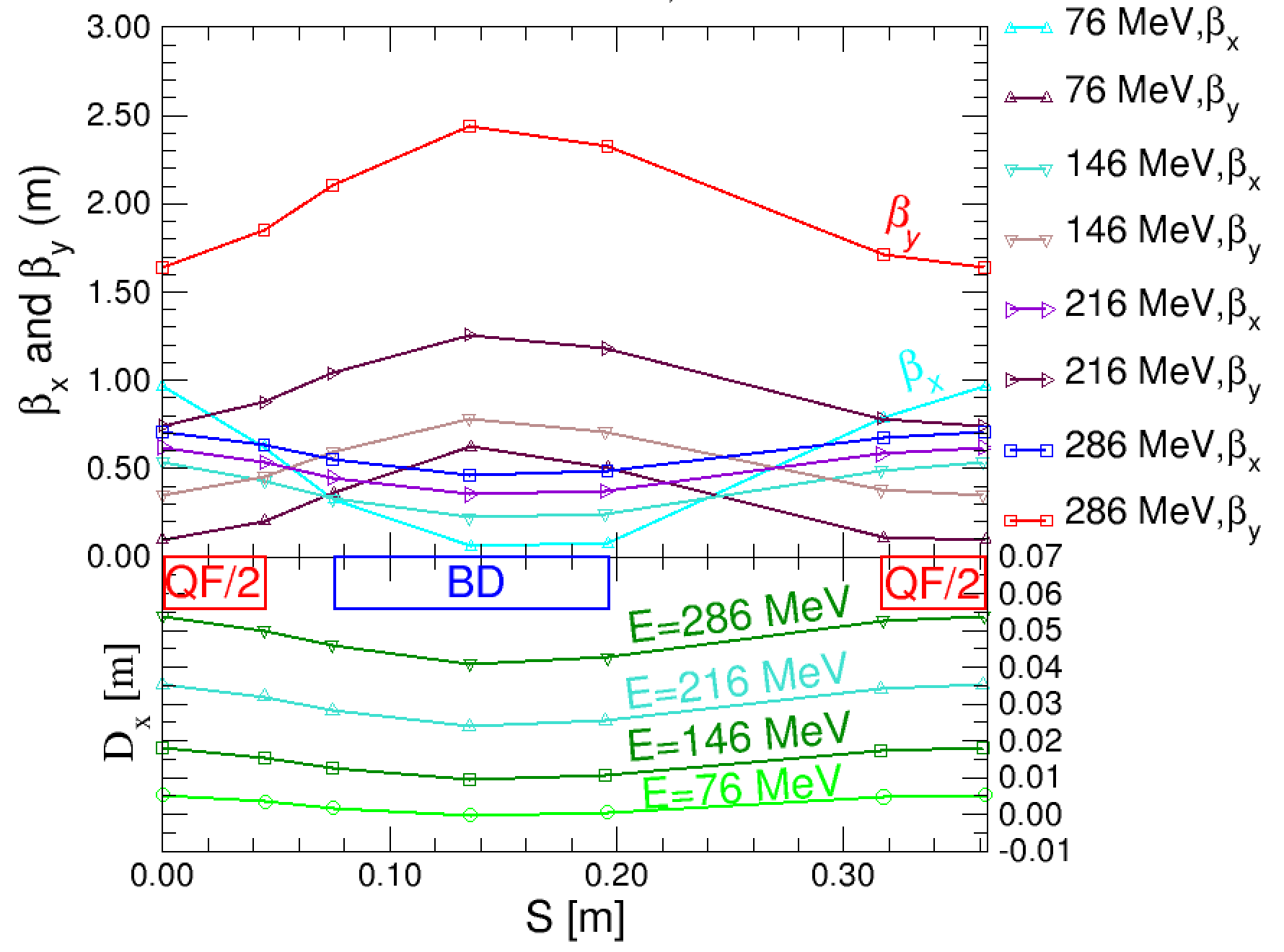}}
\caption[]{Radial offsets of the design orbits at the 4 arc energies, and matched betatron and dispersion values through the arc cell.}
\label{fig:ffag_cell_overview}
\end{figure}

\begin{figure}[tb]
\centering
\subfloat[Combined function dipole magnet]{\includegraphics[width=0.55\textwidth]{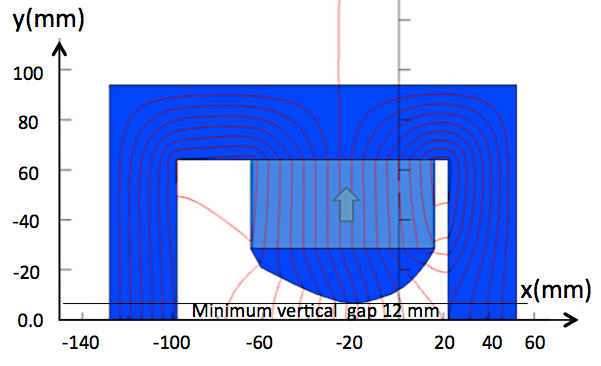}\label{fig:figure_combined_function_dipole}}
\hspace{0.05\textwidth}
\subfloat[Halbach magnet]{\includegraphics[width=0.35\textwidth]{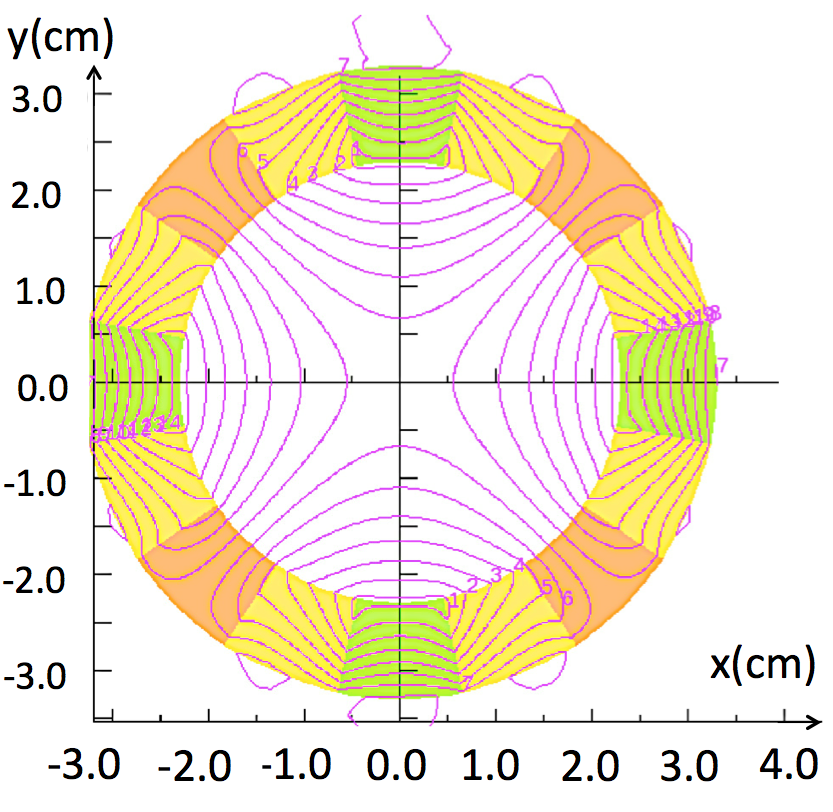}\label{fig:figure_halbach_magnet}}
\caption[]{The BD defocusing iron-dominated combined function H-type magnet, with iron in dark blue and permanent magnet material in light blue. The QF focusing Halbach permanent magnet quadrupole, with an overall diameter of about 70 mm.  There is ample room for vacuum pipe placement, despite orbit offsets in the range from -6 mm to +16 mm.  All systematic multipole errors below 12-pole are zero, thanks to the 4-fold symmetry.}
\label{fig:ffag_magnet_overview}
\end{figure}

\Figure{fig:ffag_cell_overview} shows the radial orbit offsets for the 4 arc energies, and the small values of the matched cell betatron and dispersion functions resulting from the very strong focusing, while \Fig{fig:ffag_magnet_overview} shows cross sections of the BD and QF magnets.    \Table{tab:ffag_cell_behavior} lists the tune-per-cell values and the path length differences for the 4 arc energies, while \Tab{tab:ffag_cell_parameters} shows the parameters of the arc unit cell.

\begin{table}[tb]
\caption[]{NS-FFAG arc cell behavior.  The path length through two FFAG arcs depends parabolically on the deviation of the arc energy from a reference value, because design orbits are radially offset by differing amounts from the arc magnet centers.}
\begin{tabular*}{\columnwidth}{@{\extracolsep{\fill}}lrrrr}
\toprule
Arc energy (MeV)	&	H tune & V Tune	& Path length difference (mm) \\
\midrule
76	& 0.406	& 0.282	& 11.6 \\
146	& 0.180	& 0.112	& -81.3 \\
216	& 0.126	& 0.059	& -71.1 \\
286	& 0.100	& 0.033	& 0 \\
\bottomrule
\end{tabular*}
\label{tab:ffag_cell_behavior}
\end{table}

\begin{table}[tb]
\caption[]{Lengths and angles of elements in the arc cell, defining a reference curve made of circular arcs and straight lines.}
\begin{tabular*}{\columnwidth}{@{\extracolsep{\fill}}lrrrr}
\toprule
Element	& Length (mm)	& Bend angle (mrad)	& Dipole field (T)	& Gradient (T/m)	\\
\midrule
QF &	89.4	 & 24.219	& 0.21	& -39.1 \\
Drift 1	& 30.0	 & & & \\		
BD	& 120.5	& 87.051	& -0.56	& 25.42 \\
Drift 2	& 121.4	& & & \\		
Total	& 361.3	& $-2\pi/100$ & & \\
\bottomrule
\end{tabular*}
\label{tab:ffag_cell_parameters}
\end{table}

\clearpage
\section{Beam breakup}
The beam current in a conventional linear accelerator is typically limited by the electrical power available for acceleration.  This limit is lifted in ERLs because the beam energy is fed back into the accelerating fields, but a new limit to the current is created by the beam-breakup instability (BBU).  Higher Order Modes (HOMs) excited by the beam passing --- more than once --- through the SRF cavities give rise to the recirculative beam-breakup instability, when the current is too large, and/or when there are too many passes.  This leads to large beam-trajectory oscillations and beam loss.  A bunch traversing a cavity with a small dipole-HOM field receives a transverse kick.  When this bunch returns to the same cavity it will therefore excite dipole-HOMs, thanks to the transverse offset caused by the initial transverse kick.  If it excites the dipole-HOM in phase with the prior HOM field, then the transverse kick will increase and can eventually cause the beam to be lost. 

A rigorous theory shows that BBU can become very restrictive in multi-turn ERLs, with a threshold current that decreases with increasing $N$ as $1/[N(2N-1)]$, where $N$ is the total number of accelerating passes~\cite{ref10}.   Note that $N$ can be large --- it is 16 in a recent eRHIC design.  Comparisons to theory were excellent in the experimental investigations of BBU at the JLAB FEL-ERL --- including horizontal-vertical coupling, but only with one pass~\cite{ref11}.  Multi-turn BBU has been studied at CEBAF, although not in an ERL~\cite{ref12}.  The Cb accelerator will be the first to experimentally investigate multi-turn BBU in an ERL, a vital activity for eRHIC. Detailed simulations, using the codes developed at Cornell University for the HOM parameters of the designed ERL linac cryomodule~\cite{ref13} show that the BBU threshold for the proposed ERL test is between 40 and 350~mA depending on the betatron phase advance of the recirculating arcs.  Thus, the simulated threshold is sufficiently high for the proposed ERL test loop with $N = 4$.

\clearpage
\section{Injector}
\begin{figure}[tb]
\centering
\includegraphics[width=0.95\textwidth]{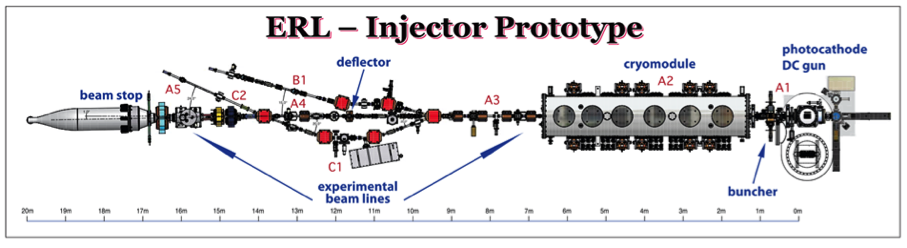}
\caption[]{The photoemission injector currently operating at Cornell accelerates the beam to 6 MeV. }
\label{fig:diagram_cornell_injector}
\end{figure}

\begin{figure}[tb]
\centering
\subfloat[Injector cryomodule]{\includegraphics[width=0.45\textwidth]{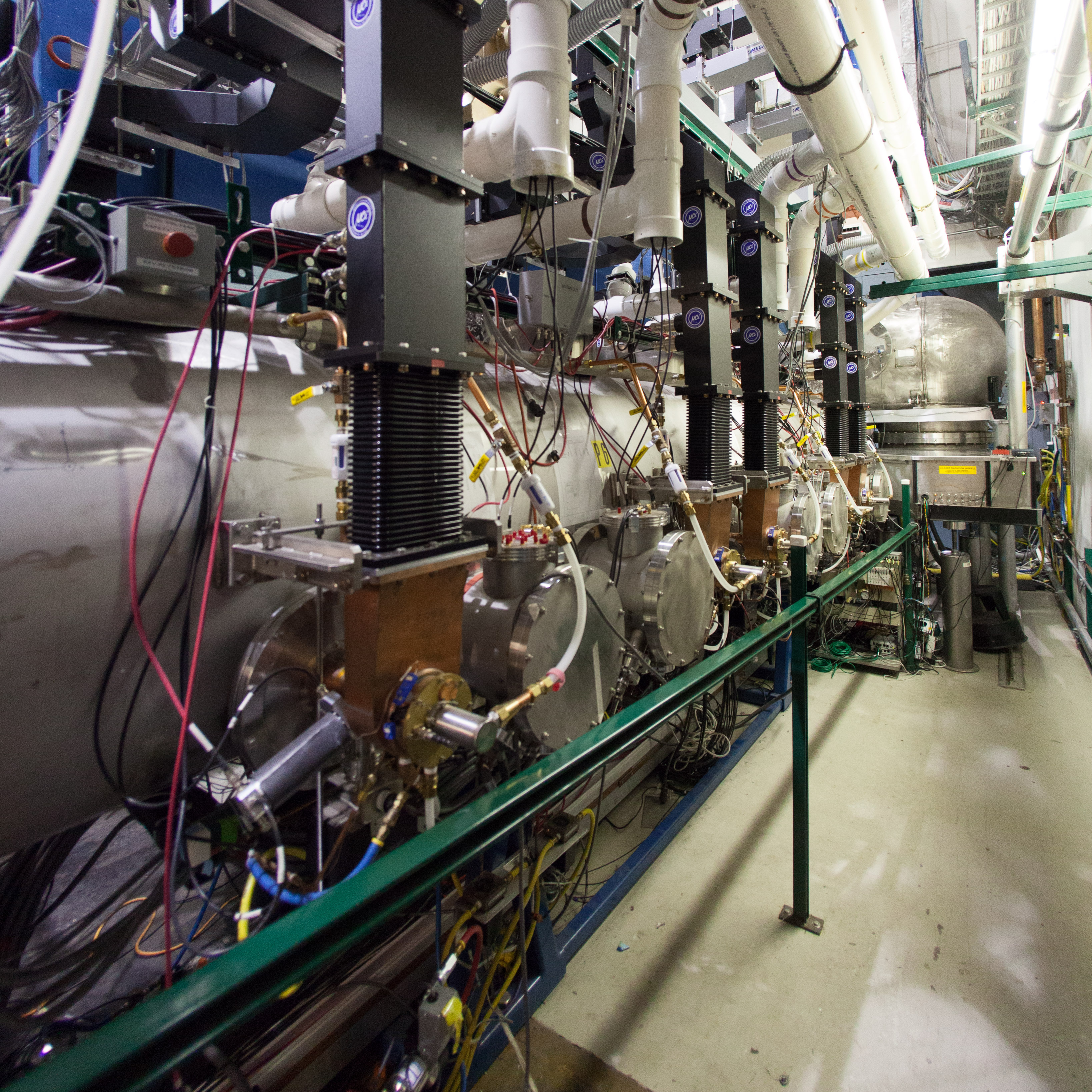}\label{fig:photo_cornell_injector}}
\hspace{0.05\textwidth}
\subfloat[DC gun]{\includegraphics[width=0.45\textwidth]{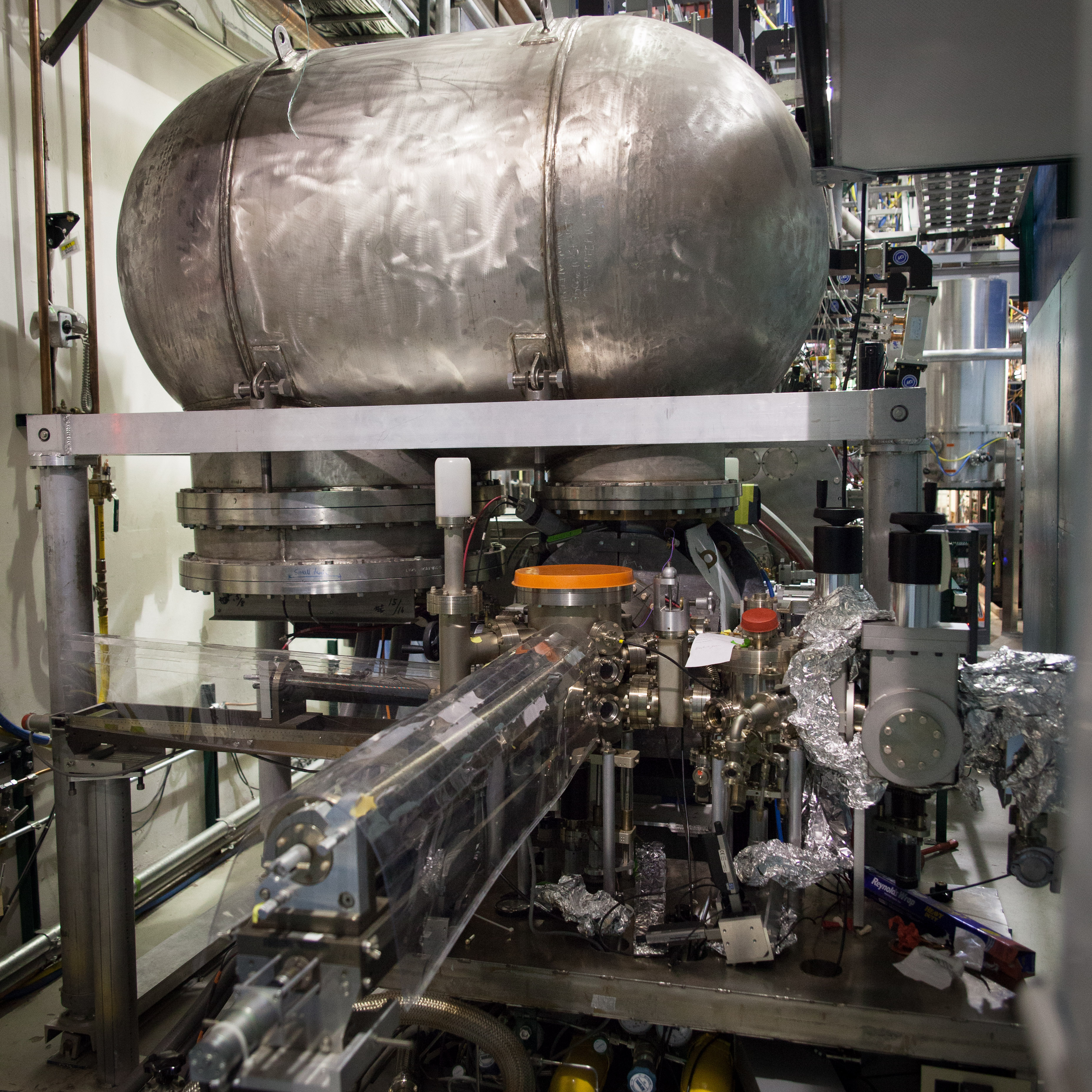}\label{fig:photo_cornell_dc_gun}}
\caption[]{ The photographs show (from right to left) the high voltage DC gun, an emittance compensation section, the RF buncher, and the cryomodule.  Accelerated beam is then directed into a beamline or into the beam stop.}
\label{fig:photos_cornell_injector}
\end{figure}

The photoemission electron injector shown in \Fig{fig:diagram_cornell_injector} and \Fig{fig:photos_cornell_injector} is fully operational, and requires no further development.   It has achieved the world-record current of 75~mA~\cite{ref14,ref15,ref16}, and record low beam emittances for any CW photoinjector~\cite{ref17}, with an equivalent brightness that outperforms the best existing storage rings by a substantial factor, if the injected beam were accelerated to a similar energy.  Cornell has established a world-leading effort in photoinjector source development, in the underlying beam theory and simulations, with expertise in guns, photocathodes, and lasers.  The strength of the injector group is in combining various cathode advances and innovations with the world's brightest photoinjector.  This ability to both formulate the frontier questions for high brightness source development, and also to implement the solutions, makes breakthroughs in accelerator science possible.

\section{SRF cryomodule}

\begin{figure}[tb]
\centering
\includegraphics[width=0.95\textwidth]{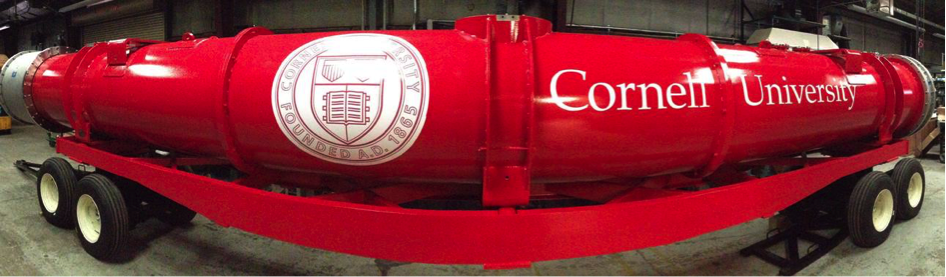}
\caption[]{The full-scale prototype 1300 MHz cryomodule that is available for C$\beta$ use.}
\label{fig:photo_cornell_main_linac_cryomodule}
\end{figure}

\Figure{fig:photo_cornell_main_linac_cryomodule} shows the full-scale prototype cryomodule that was developed and constructed as part of the  It consists of six 7-cell 1300 MHz SRF cavities, including HOM absorbers and RF power couplers.  The cavity geometry is carefully designed to maintain high Q while maximizing the beam breakup threshold current~\cite{ref18}.  Cryomodule construction is complete, and the device is available for use in this project.  The cryogenics necessary to cool the linac and injector cryomodules are available from the Wilson Lab cryoplant.  Additional hardware will be required to transfer the cryogens from the cryoplant to the cryomodules, including heat exchanger cans, valve boxes and 2~K pump skids.

\section{RF power}

The injector delivers up to 500 kW of RF power to the beam at 1300~MHz.  The buncher cavity uses a 16~kW IOT tube, which has adequate overhead for all modes of operation.  The injector cryomodule is powered through ten 50 kW input couplers, using five 130 kW CW klystrons.  The power from each klystron is split to feed two input couplers attached to one individual 2-cell SRF cavity.  An additional klystron is available as a backup, or to power a deflection cavity for bunch length measurements.  The main linac cryomodule will be powered by 6 individual solid-state RF amplifiers with 5 kW average power per amplifier.  Each cavity has one input coupler.  One amplifier is currently available for testing purposes, so an additional 5 amplifiers are needed for this project.

\clearpage
\section{Beam instrumentation and bunch patterns}

Beam instrumentation will support three operational modes:
\begin{enumerate}
  \item Initial commissioning: one bunch with about 0.1 nC/bunch..
  \item Physics operation: CW-like beams, for example with one accelerating (and one decelerating) low charge bunch in every RF period.
  \item Dedicated experiment: eRHIC-like trains, typically with bunches spaced by more than 2~ns (3 RF buckets), and no more than 5 nC per bunch.
\end{enumerate}

\begin{figure}[tb]
\centering
\includegraphics[width=0.95\textwidth]{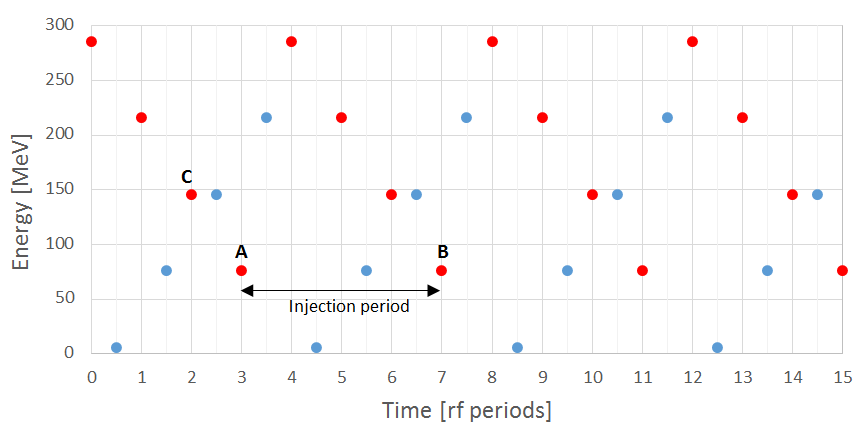}
\caption[]{Bunches passing a reference point at the end of the linac, in {\bf physics operation} mode with one accelerating bunch (and one decelerating bunch) in every RF period.  Red and blue dots represent accelerating and decelerating bunches, respectively.  Bunch B, injected 4 periods after bunch A, also has an energy of 76 MeV.  Bunch C was injected 280 periods before bunch  A , and so has an energy of 146 MeV because it has passed through the linac twice.}
\label{fig:plot_bunch_pattern}
\end{figure}

\Figure{fig:plot_bunch_pattern} illustrates bunch pattern behavior in Physics operation mode, observing bunches as they pass a reference point at the end of the linac.  Since each bunch is accelerated four times and decelerated four times, the maximum injection rate (that avoids overlapping bunches) is one quarter of the RF frequency.  Overlap avoidance also requires the revolution harmonic, 279, to be an odd number.  The maximum energy turn is shorter (or longer) than the other turns by half a wavelength, returning each bunch with the correct phase for deceleration.

The instrumentation design and commissioning strategy for CW, single-pass ERL operation is documented in~\cite{ref19}.   With multi-pass ERL operation, a time-isolated `diagnostic' bunch needs to be monitored.  Bunch properties --- including beam intensity and loss, position, energy, phase, transverse and longitudinal beam sizes --- will be measured at different energies in the splitter-combiners, experimental straights, and in the beam stop.  Optical function measurements enable beam emittance determinations.  The broad range of beam parameters in the different operating modes requires careful study to determine if different hardware and/or signal processing electronics will be required. 

Instrumentation prototyping goals motivated by eRHIC include developing and demonstrating:
\begin{enumerate}
  \item  A high time resolution BPM design (BNL R\&D is ongoing in FY15/16).
  \item  Ion-clearing mechanisms to eliminate conventional ion trapping and fast ion instabilities.
  \item  High-bandwidth diagnostics for resolving beam sizes in the FFAG cells.
\end{enumerate}

\section{Control system}
The accelerator will use EPICS~\cite{ref20} in order to minimize development time, ensure scalability, and avoid performance limitations.  This choice enables the use of hardware and software already developed at Cornell and at BNL.  It also provides access to well-defined interfaces at both the server and client levels that are used at many accelerator facilities, since EPICS has a large user base in the accelerator community.  EPICS also has well defined developer interfaces and infrastructure allowing rapid development for new systems. 

The control system enables supervisory control, automation, and operational analysis, with a scope that extends from the equipment interface to accelerator operators, and to experimenters and technical staff.  It includes global systems such as timing, deterministic data communication, network communication, control room operations, automation and optimization, in addition to the computers and software required to implement and integrate all subsystems.  The topmost layer of the client-server architecture provides access for non-real time activities, for example high level physics modeling using live or stored data.  Accelerator operation and monitoring take place in a middle layer.  A dedicated equipment layer, lower down, interfaces to specific equipment through point-to-point protocols.  Other low level layers synchronize sub-systems (such as vacuum controls and personnel protection systems) with the accelerator at lower rates.

\section{Conclusion}
We have described a proposed collaborative effort between Brookhaven National Laboratory and Cornell University to design, build, and test a new type of energy recovery linac.  

The FFAG-ERL has the potential to deliver excellent beam quality at a cost that is reasonable, compared to the current state-of-the-art, for a wide range of physics and applied physics applications.

This collaborative effort would be a model for future joint national laboratory/university projects, taking advantage of the expertise and skills that both offer, to develop exciting new technologies in a timely and cost-efficient manner.

This FFAG-ERL prototypes important features of the BNL eRHIC project.

All major technical components except for the FFAG arcs already exist at Cornell University, and have mostly been commissioned.  Suitable space exists at Cornell, and planning for its refurbishment has already started.


\end{document}